\documentclass{ws-mpla}

\usepackage{amsmath}
\usepackage{amssymb}
\usepackage{graphicx}

\begin{document}

\markboth{Liu \textit{et al}}
{A modified model of  non-locally corrected  gravity: cosmology and Newtonian limit}

\catchline{}{}{}{}{}

\title{A modified model of  non-locally corrected  gravity: \\
Cosmology and the Newtonian limit}

\author{\footnotesize Dao-Jun Liu}

\address{Center for Astrophysics, Shanghai Normal University,\\
 100 Guilin Road, Shanghai 200234,China\\
djliu@shnu.edu.cn}

\author{Bin Yang}
\address{Center for Astrophysics, Shanghai Normal University,\\
 100 Guilin Road, Shanghai 200234,China}

\author{Xing-Hua Jin}
\address{Department of Mathematics, Shanghai Business School, Shanghai 200235, China}

\maketitle

\pub{Received (Day Month Year)}{Revised (Day Month Year)}

\begin{abstract}
A modified form of non-locally corrected theory of gravity is investigated in the context of cosmology and the Newtonian limit. This form of non-local correction to classic  Einstein-Hilbert action can be locally represented by a triple-scalar-tensor theory in which one of the scalar degree of freedom is phantom-like and the other two are quintessence-like. We show that there exists a stable de Sitter solution for the cosmological dynamics if a suitable form of potential function $V(\phi)$ (or equivalently, $f(R)$) is selected. However, no matter what a potential function  is selected, there is always an early time repeller solution corresponding to a radiation dominated universe.
Besides, the equations for linear scalar perturbations are presented and it is shown that the form of potential function $V(\phi)$ is stringently constrained by the Solar system test, although the post-Newtonian parameter $\gamma$ is not directly affected by this function.  
\end{abstract}

\ccode{PACS Nos.:{98.80.-k, 95.36.+x, 04.50.-h}}

\section{Introduction}

Strong evidences from the current cosmological observations  converge upon the fact that the universe is spatially flat and there exists exotic component, dubbed dark energy, which drives the speed-up expansion of the universe. Many scenarios have been proposed to explain the acceleration in the framework of general relativity (GR). The preferred and simplest candidate for dark energy is the Einstein's cosmological constant which can fit the observations well. However, it suffers from the so-called fine-tuning problem  and coincidence problem. Meanwhile, the observations are not yet able to confirm that dark energy is indeed a constant. Actually,  many dynamical models of dark energy have been studied extensively during the past over ten years, for a comprehensive review, see for example \cite{Copeland:2006ej,Li2011}. 

As is pointed out, the dark energy problem may be essentially an issue of quantum gravity \cite{Witten:2002}. Although a complete  theory of quantum gravity has not been established, some valuable ideas are thought to be the features of the theory of quantum gravity, such as non-locality.  Non-locality can naturally arise in string theory and the non-local corrections to the  effective action of gravity can be generated from quantum loop effects.  It is also believed that the black hole information paradox may be  resolved by the effect of non-locality \cite{Giddings2006}. Recently, a new non-local corrected theory of gravity  was suggested in the form
\begin{equation}\label{action0}
S=\frac{1}{2\kappa}\int d^4xR(1+f(\Box^{-1}R)),
\end{equation}
where $\Box^{-1}R$ denotes the inverse d'Alembertian acting on the scalar curvature\cite{Deser2007}.  In the limit of $f=0$, it goes back to general relativity.  It was found that this kind of non-local gravity may lead to a unified description of the early-time universe  and the present accelerating universe\cite{Deser2007,Nojiri2008,Koivisto2008} and there is no apparent conflicts with observations in the solar system tests\cite{Koivisto2008b}.  Non-local gravity has been extensively investigated in the literature \cite{Nonlocal,Nonlocal2,Nonlocal3,Nonlocal4,Nonlocal5,Nonlocal6,Nonlocal7,Calcagni2010,Nonlocal8,Nonlocal9}.

 In this work, we shall suggest a modified non-local term to the effective action of gravity and investigate its implications in cosmology and the solar system tests.  Although,  just as that  in action (\ref{action0}),  this form of non-local corrections to gravity has also no direct support from fundamental theories, we wish it appears to provide a small window through which one can look into phenomenological aspects of fundamental theories.

 This paper is organized as follows:  we first present the model in general form and a triple-scalar tensor representation is given in Sec. \ref{sec:formalism}. In Sec. \ref{sec:cosmology}, we investigate the cosmological dynamics of the theory. In Sec. \ref{sec-1} the scalar mode  of cosmological perturbation  to the first order  and its Newtionian limit are considered. Finally, in section \ref{sec:conclusion}  we summarize our results and give some discussions.

\section{Non-local $f(R)$ gravity and triple-scalar-tensor theory}
\label{sec:formalism}
 We here consider the following simple modified action of the non-locally corrected gravity 
  \begin{eqnarray}
    \label{eq:OrginAction}
    S=\int d^4x\sqrt{-g}\left\{\frac{1}{2\kappa}[R(1+\Box^{-1}f(R))]+\mathcal{L}_{m}\right\},
  \end{eqnarray}
where $g$ is the determinant of metric tensor $g_{\mu\nu}$, $\kappa\equiv 8\pi G$ and  $f$ is a generic function which has the same dimension with Ricci scalar $R$. Here, $\Box^{-1}$ is the inverse d'Almbertian operator, which is in fact an integral operator,  therefore, the above action is non-local. The equations of motion derived directly from this action will be rather difficult to handle because of the non-locality.  However, by introducing two Lagrange multipliers $\eta$ and $\xi$ and replacing the inverse d'Alembertian and Ricci scalar as two  auxiliary scalar fields $\psi$ and $\varphi$ respectively,  the above action can be rewritten in the following form,
  \begin{eqnarray}
    \label{eq:Action2}
    S = \int
    d^4x\sqrt{-g}&&\left\{\frac{1}{2\kappa}\{\varphi(1+\psi)+\eta(R-\varphi)\right.\nonumber\\
   &&\left. -\xi(\Box \psi-f(\varphi))\}+\mathcal{L}_{m}\right\}. 
  \end{eqnarray}
It can be checked, by varying action (\ref{eq:Action2}) with respect to $\eta$, $\xi$ and $\varphi$ respectively, that  $R=\varphi$, $\Box\psi=f(\varphi)$ as we expect and $\eta=1+\psi+\xi f'(\varphi)$.
Integrating partially and neglecting the boundary terms, then the double derivative is eliminated as usual, we obtain a kind of triple-scalar-tensor theory
\begin{eqnarray}
\label{eq:action3}
S=\int d^4x\sqrt{-g}&&\left\{\frac{1}{2\kappa}\left[ R(1+\psi+\xi f'(\varphi))-\varphi\xi f'(\varphi)\right.\right.\nonumber\\
&&\left.\left.+\partial^\lambda  \xi\partial_{\lambda} \psi+\xi f(\psi)  \right] +\mathcal{L}_{m}\right\}.
\end{eqnarray}   
 It is noted that this kinetic coupling can not eliminated by a field redefinition. To be convenient, defining $\phi= f'(\varphi)$ and taking $\varphi$ as a function of $\phi$, then  introducing a new function 
 $V(\phi)= \varphi(\phi)\phi-f(\varphi(\phi))$, we can rewritten action (\ref{eq:action3}) as
  \begin{eqnarray}
  \label{eq:action4}
  S&=&\int d^4x\sqrt{-g}\left\{\frac{1}{2\kappa}\left[ R(1+\psi+\xi\phi)\right.\right.\nonumber\\
  &&\left.\left.-\xi V(\phi)+\partial^\lambda  \xi\partial_{\lambda} \psi  \right] +\mathcal{L}_{m}\right\}.
  \end{eqnarray}
Contrary to the case for the action (\ref{action0}), in which the local equations of motion can be derived by varying with respect to the metric and two auxiliary fields, there are three auxiliary local fields in our model  (\ref{eq:action4}). 

By varying action(\ref{eq:action4}) with respect to the metric $g_{\mu\nu}$,  we obtain the following modified Einstein equations
\begin{eqnarray}
  \label{eq:Einstein1}
&&FG_{\mu\nu}+\frac{1}{2}g_{\mu\nu}(\xi V(\phi)-\partial^{\lambda}\xi\partial_{\lambda}\psi)\nonumber\\
&& +\partial_{\mu}\xi\partial_{\nu}\psi+(g_{\mu\nu}\Box-\nabla_{\mu}\nabla_{\nu})F
=\kappa T_{\mu\nu},
\end{eqnarray}
where we have defined $F=1+\psi +\xi\phi$ and $G_{\mu\nu}$ is the Einstein tensor defined as usual and $T_{\mu\nu}$ denotes the stress tensor of  matter.
Furthermore,  the variation with respect to $\psi$, $\xi$ and $\phi$ gives the following equations of motion
\begin{equation}
\Box\psi=V'(\phi)\phi-V(\phi),
\end{equation}
\begin{equation}
  \label{eq:4}
  \Box\xi=V'(\phi)
\end{equation}
and $\xi(R-V'(\phi))=0$. When $\xi$ is not zero, we have
\begin{equation}
\label{eq:RVprime}
R=V'(\phi),
\end{equation}
where the prime denotes the derivative with respect to $\phi$.
 Because matter is minimally coupled with other fields, the matter energy-momentum tensor is conserved, $\nabla^\mu T^m_{\mu\nu}=0$. Note that this conservation law is not an independent condition.

\section{Cosmological dynamics}
\label{sec:cosmology}
In this section, we shall investigate the cosmological dynamics of the non-locally corrected theory presented in the above section.   In a flat 4-dimensional  Friedmann-Robertson-Walker(FRW) space-time, 
\begin{equation}
  \label{eq:5}
  ds^2=-dt^2+a(t)^2\left[dr^{2}+r^{2}d\theta^{2}+r^{2}\sin^{2}\theta
  d\phi^{2}\right],
\end{equation}
we obtain the Friedmann equations 
\begin{equation}\label{eq:friedman}
3FH^2-\frac{1}{2}(\xi V(\phi)-\dot{\xi}\dot{\psi})+3\dot{F}H=\kappa\rho_m,
\end{equation}
\begin{equation}\label{eq:ddotF}
\ddot{F}+2H\dot{F}+ (2\dot{H}+3H^2)F-\frac{1}{2}(\xi V(\phi)+\dot{\xi}\dot{\psi})=-\kappa p_m
\end{equation}
where
 $H = \dot{a}/a$ is the Hubble parameter, the overdots denote the derivatives with respect to cosmic time $t$,  $\rho_m$ and $P_m$ are energy density and pressure of matter, respectively.

The two scalar field equations become 
\begin{equation}
  \label{eq:8}
  \ddot{\psi}+3H\dot{\psi}+V'(\phi)\phi-V(\phi)=0,
\end{equation}
\begin{equation}\label{eq:ddxi}
\ddot{\xi}+3H\dot{\xi}+V'(\phi)=0.
\end{equation}
Besides, the usual matter conservation law reads
\begin{equation}
  \label{eq:22}
  \dot{\rho_m}+3H(1+w)\rho_m=0,
\end{equation}
where we have assumed an equation of state for the matter $p_m=w\rho_m$. It is worth noting that, due to the
Bianchi identity, only four equations among Eqs.(\ref{eq:friedman}-\ref{eq:22}) are independent.

\begin{figure}[hbtp]
\centering
\includegraphics[width=0.9\linewidth]{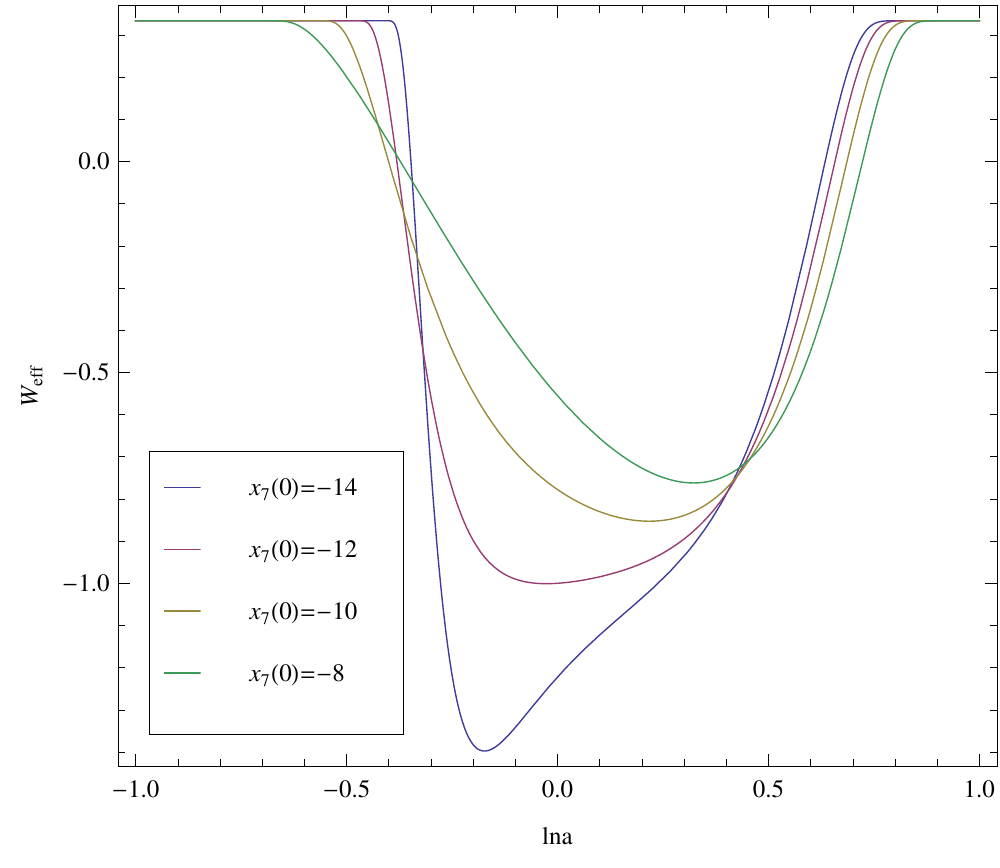}
\caption{The  evolution of the  effect equation of state $w_{eff}$ with respect to $\ln a$ in the exponential model $V(\phi)=V_0e^{\alpha \phi}$ for different initial conditions , where the parameter $\alpha$ is set to be $-1$.  }
\label{fig1}
\end{figure}

To rewrite the equation of motions as a dynamical system,  we may define new dimensionless variables
\[\Omega_m=\frac{\kappa \rho_m}{3H^2},\;\; x_1=\xi\phi,\;\; x_2= \frac{\dot{\xi\phi}}{H}, \;\; x_3=\xi,\;\; x_4= \frac{\dot{\xi}}{H}, \]
\[x_5=\psi,\;\; x_6=\frac{\dot{\psi}}{H},\;\; x_7=\frac{V(\phi)}{H^2},\;\; x_8=-\frac{V'(\phi)}{V(\phi)}  \]
as the functions of $N=\ln a$.  Then, the cosmological dynamics can be represented by the following equations
\begin{equation}
x_1'=x_2,
\end{equation}\label{eq:dx2}
\begin{eqnarray}
x_2'&=&\frac{1}{3}\left[2x_3x_7+x_4x_6+x_7x_8(1+x_1+x_5-3\frac{x_1}{x_3})\right]\nonumber\\
&&+(\frac{1}{6}x_7x_8-1)x_2-x_7-(3w-1)\Omega_m,
\end{eqnarray}
\begin{equation}
x_3'=x_4,
\end{equation}
\begin{equation}\label{eq:dx4}
x_4'=(\frac{1}{6}x_7x_8-1)x_4+x_7x_8,
\end{equation}
\begin{equation}
x_5'=x_6,
\end{equation}
\begin{equation}\label{eq:dx6}
x_6'=(\frac{1}{6}x_7x_8-1)x_6+(\frac{x_1x_8}{x_3}+1)x_7,
\end{equation}
\begin{equation}\label{eq:x7}
x_7'=(\frac{1}{3}x_7x_8+4)x_7-\frac{x_7x_8}{x_3^2}(x_2x_3-x_1x_4),
\end{equation}
\begin{equation}\label{eq:x8}
x_8'=-(\Gamma-1)\frac{x_8^2}{x_3^2}(x_2x_3-x_1x_4),
\end{equation}
\begin{equation}
\Omega_m'=\Omega_m\left[\frac{1}{3}x_7x_8+4-3(1+w)\right],
\end{equation}
where the primes denote the derivative with respect to $N$ and the parameter $\Gamma$ is defined to be $\Gamma={VV''}/{V'^2}$.
The Friedman equation is now a constraint equation
\begin{equation}
\Omega_m=1+x_1+x_2+x_5+x_6+\frac{1}{6}x_4x_6-\frac{1}{6}{x_3}{x_7}.
\end{equation}
For an exponential potential $V(\phi)=V_0e^{\alpha\phi}$, $\Gamma=1$. Then, from Eq.(\ref{eq:x8}), we know $x_8=-\alpha$ is a constant.  One the other hand, it is clear that $ x_7=0 $ is a solution of Eq.(\ref{eq:x7}). Therefore, we can easily obtain the following solution
\[ \Omega_m=\Omega_m^0e^{(1-3w)N}, \]
\[ x_6=C_6e^{-N}, \]
\[ x_5=C_5-C_6 e^{-N}, \]
\[ x_4=C_4e^{-N}, \]
\[ x_3=C_3-C_4e^{-N}, \]
\[ x_2=C_2e^{-N} +  \frac{(3 w-1) \Omega_m^0 }{3 w-2}e^{(1-3 w)N}-\frac{1}{3} C_4 C_6 e^{-2N}, \]
\[ x_1=C_1- C_2e^{-N} -\frac{ \Omega_m^{0} }{3 w-2}e^{ (1-3 w)N}+\frac{1}{6}C_4 C_6e^{-2N}, \]
where $C_i(i=1..6)$ and $\Omega_m^0$ are constants. The Friedman equation requires that  $1+C_1+C_5=0$. 
From $x_7=0$, we know in this solution the scalar curvature $R=0$, hence, $a=a_0\sqrt{1-2H_0t_0+2H_0t}$, where we have set $H(t_0)=H_0$ and $a(t_0)=a_0$. Clearly, when the equation of state $w>1/3$, the solution is stable and the fixed point $(x_1=C_1, x_2=0, x_3=C_3, x_4=0, x_5=C_5=-C_1-1, x_6=0, x_7=0)$ is an atrractor.  In figure \ref{fig1}, we show the evolution of effect equation of state   $w_{eff}=-\frac{2\dot{H}}{3H^2}-1$ in the exponential model for different initial conditions. It is found that there exist a stable attractor where $w_{eff}=1/3$. We note that in this model, the acceleration of the universe depend heavily on the ininital conditions.  This solution may be a good representation of the early universe when a radiation dominated phase follows naturally from an inflationary expansion.  Also, it is interesting to find that there is a case with realizing a crossing of the phantom divide from the
non-phantom (quintessence) phase to the phantom one for  some time and then going back to the non-phantom phase at  late time.  

\begin{figure}[hbtp]
\centering
\includegraphics[width=\linewidth]{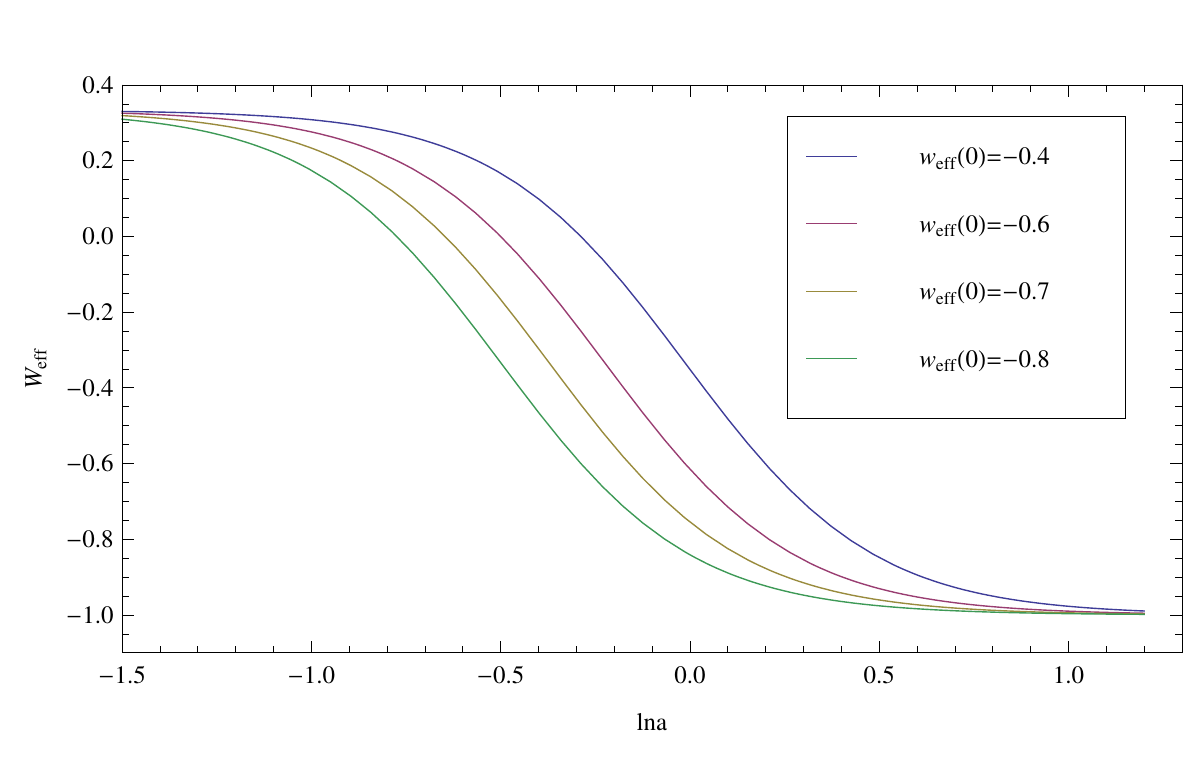}
\caption{The  evolution of the  effect equation of state $w_{eff}$ with respect to $\ln a$ in the linear model $V(\phi)=V_0\phi$ for different initial conditions.  }
\label{fig2}
\end{figure}

Let us now consider whether there is  a de Sitter solution of the theory in which  the Hubble parameter is a constant,  $H(t)=H_0$ and the scale factor $a=a_0 e^{H_0 t}$. We assume that $\xi$ is not zero,  so $V'(\phi)$ in the equations is just the scalar curvature $R$, which  is also a constant in this case, $R=12H_0^2$.   Therefore, from Eq.(\ref{eq:ddxi}),  the solution for $\xi$ can be straightforward obtained,
\begin{equation}
\xi=C_2-4H_0t-\frac{C_1}{3H_0}e^{-3H_0t}.
\end{equation}
For simplicity, we let the intergal constants $C_1$ and $C_2$ vanish. This means $x_3=-4N$.
We also assume the potential function $V(\phi)$ is given by a linear function  $V(\phi)=V_0\phi$, where the constant $V_0$ should be equal to $12H_0^2$ due to the relation (\ref{eq:RVprime}). Therefore, from the definition of  $x_7$ and $x_8$, we have $x_7x_8=-12$.  Applying this relation and assuming $x_2=0$, we obtain that $x_1=A$ is a constant, $x_4=-4$, $x_5=A-1-\frac{1}{3}Be^{-3N}$, $x_6=Be^{-3N}$, $x_7=-3A/N$, $x_8=4N/A$, $\Omega_m=\Omega_m^0e^{-3(1+w)N}$, where $A$, $B$ and $\Omega_m^0$ are three constants. Thus,  using cosmic time $t$,  we can represent the solution as $\phi=-\frac{A}{4H_0 t}$,
$F=2A-\frac{B}{3}e^{-3H_0 t}$ and $\psi=A-1-\frac{B}{3}e^{-3H_0 t}$.  Note that if the potential $V$ is a linear function of $\phi$,  from its definition,  $f(R)$ should be also a linear function, that is, the non-locally corrected term in the action reads 
$\sim R\Box^{-1}(R-V_0)$. In Fig. \ref{fig2}, we plot the trajectories of the evolution of $w_{eff}$ under different initial conditions.  It is shown that $w_{eff}$  trends to the limit $-1$ in the future regardless of what the initial value takes, which may means that the de Sitter solution above is a stable  attractor solution.  Meanwhile, in the past, there may exists an repeller solution  representing an effective radiation dominated  universe in which $w_{eff}=1/3$.

\section{Scalar perturbations and Newtonian limit}\label{sec-1}
In this section, we show cosmological perturbation equations in our theory. Clearly, the evolution of scalar perturbations are much more complicated than those of the vector and tensor perturbations because they are  heavily  coupled to scalar fields and matter.  Here we only consider scalar perturbations.
As usual, once the scalar modes of perturbations are introduced, the perturbed metric can be written in the Newtonian gauge to the first order  as
\begin{eqnarray}
\label{eq:10}
ds^2=-(1+2\Psi(\vec{x},t))dt^2+a(t)^{2}\delta_{ij}(1+2\Phi(\vec{x},t))dx^idx^j,
\end{eqnarray}
where $\Psi$ and $\Phi$ are two scalar potentials. When both of the two potentials vanish, the FRW metric of the background universe is recovered. 
Meanwhile, the three scalar fields are perturbed to be $\psi(\vec{x},t)=\psi(t)+\delta\psi(\vec{x},t)$,  $\xi(\vec{x},t)=\xi(t)+\delta\xi(\vec{x},t)$ and $\phi(\vec{x},t)=\phi(t)+\delta\phi(\vec{x},t)$ respectively. 
Similarly,  as for matter, the energy density and pressure are  perturbed to be $\rho_{m}(\vec{x}, t) =\rho_{m}(t)+\delta\rho_{m}(\vec{x}, t)$ and $P_{m}(\vec{x}, t)=P_{m}( t)+\delta P_{m}(\vec{x}, t)$.   Inserting these expressions into Einstein equation (\ref{eq:Einstein1}), up to the first order, we obtain the following equations for  the perturbations 
\begin{eqnarray}
\label{eq:40}
& &2F\left[3H^2{\Psi}-3H\dot{\Phi}+a^{-2}\nabla^{2}\Phi\right]-3H^{2}\delta F\nonumber\\ 
& &+\frac{1}{2}\left[\xi
R\delta\phi+V(\phi)\delta\xi-3(\dot{\psi}\dot{\delta\xi}+\dot{\xi}\dot{\delta\psi}+2\Psi\dot{\xi}\dot{\psi})\right]\nonumber\\
& &+3H(2\Psi \dot{F}-\dot{\delta F})-3\dot{\Phi}\dot{F}+a^{-2}\nabla^{2}\delta
F=-\kappa\delta\rho_{m},
\end{eqnarray}
\begin{eqnarray}
\label{eq:50}
&&2F(\Phi_{,0i}-H\Psi_{,i})-\left[\dot{\xi}(\delta\psi)_{,i}+\dot{\psi}(\delta\xi)_{,i}\right]\nonumber\\
&&+(\delta F)_{,0i}-\dot{F}\Psi_{,i}-H(\delta F)_{,i}=0
\end{eqnarray}
and 
\begin{eqnarray}\label{eq:70}
&\delta^{i}_{\ j}& \left\{\right.(2\dot{H}+3H^{2})(2F\Psi-\delta
  F)+\frac{1}{2}\left[V(\phi)\delta\xi+\xi R\delta\phi\right]-2F\ddot{\Phi} \nonumber\\
&+&(\dot{\Psi}-3\dot{\Phi})(2HF+\dot{F})+2\Psi(\ddot{F}+2H\dot{F})-(\ddot{\delta
  F}+2H\dot{\delta F})\nonumber\\
&+& a^{-2}\left[F(\nabla^{2}\Phi+\nabla^{2}\Psi)+\nabla^{2}\delta
    F\right]\left.\right\}\nonumber\\
&-&F(\Psi_{,ij}+\Phi_{,ij})-a^{-2}\delta F_{,ij}=\kappa\delta^{i}_{\ j}\delta P_{m},
\end{eqnarray}
where $\delta F=\delta\psi+\phi\delta\xi+\xi\delta\phi$.
Similarly, from equations of motion for the scalar fields, we have
\begin{eqnarray}
\label{eq:80}
 &-&(\ddot{\delta\xi}+3H\dot{\delta\xi})+\dot{\xi}(\dot{\Psi}-3\dot{\Phi})-6\left[\ddot{\Phi}-H(\dot{\Psi}-4\dot{\Phi})\right]\nonumber \\
 &+&2\Psi\left[(\ddot{\xi}+3H\dot{\xi})+6(\dot{H}+2H^{2})\right]\nonumber \\
  &+&a^{-2}\left[\nabla^{2}\delta\xi+2(\nabla^{2}\Psi+2\nabla^{2}\Phi)\right]=0,
\end{eqnarray}
and
\begin{eqnarray}
\label{eq:90}
&-&(\ddot{\delta\psi}+3H\dot{\delta\psi})+\dot{\psi}(\dot{\Psi}-3\dot{\Phi}) -6\phi\left[\ddot{\Phi}-H(\dot{\Psi}-4\dot{\Phi})\right]\nonumber \\
&+&2\Psi\left[(\ddot{\psi}+3H\dot{\psi})+6\phi(\dot{H}+2H^{2})\right]\nonumber \\
&+&a^{-2}\left[\nabla^{2}\delta\psi+2\phi(\nabla^{2}\Psi+2\nabla^{2}\Phi)\right]=0.
\end{eqnarray}

In the Newtonian limit of the cosmological perturbations, we may omit the cosmological expansion and set scale factor $a(t)=1$.  Moreover, the derivatives with respect to time, relative to those with respect to space, are also negligibly small.  Therefore, in the vacuum outside a mass where the source term can be neglected,  the shear constraint (\ref{eq:70}) gives $$\frac{1}{3}\delta^{i}_{\ j}\left[F(\nabla^{2}\Psi+\nabla^{2}\Phi)+\nabla^{2}\delta
F\right] =\left[F(\Psi_{,ij}+\Phi_{,ij})+\delta F_{,ij}\right],$$ which shows that the gravitational potentials are not equal,
\begin{equation}
\Psi+\Phi=-\frac{\delta F}{F}.
\end{equation}
Meanwhile,  from the perturbation equations (\ref{eq:80}) and (\ref{eq:90}), we have
\begin{eqnarray}
\label{eq:110}
 \nabla^{2}\delta\xi+2(\nabla^{2}\Psi+2\nabla^{2}\Phi)=0,\\
\nabla^{2}\delta\psi+2\phi(\nabla^{2}\Psi+2\nabla^{2}\Phi)=0.
\end{eqnarray}
Solving the above two equations and using the condition that the usual Schwarzschild solution is recovered in the limit of general relativity, we obtain that
\begin{equation}
\Psi+2\Phi=-\frac{\delta\xi}{2}=-\frac{\delta\psi}{2\phi}.
\end{equation}
The perturbation $\delta\phi$ may be neglected  because field $\phi$ is not dynamical. Therefore, 
applying the relation $\delta F=\delta\psi+\phi\delta\xi$ to eliminating  $\delta F$, $\delta \psi$ and $\delta \phi$, we get the post-Newtonian parameter
\begin{equation}
\gamma \equiv -\frac{\Phi}{\Psi}=\frac{F-4\phi}{F-8\phi},
\end{equation}
which is stringently constrained to its value in general relativity $\gamma=1$  from the solar system tests such as the Cassini spacecraft \cite{Will2005}. It should be noted that $\gamma$ is not directly affected by the potential function $V(\phi)$ (or equivalently, $f(R)$), the constraints is still imposed because the value of $F$ and $\phi$ are heavily dependent on $V(\phi)$.  

\section{Conclusions}
\label{sec:conclusion}
In this work, we investigated a new form of non-locally modified theory of gravity with action (\ref{eq:OrginAction}). Just as action (\ref{action0}) can be recast as a local biscalar-tensor theory, this non-local theory can be locally represented by a triple-scalar-tensor theory. 
In the context of cosmology, we show that there exists a stable de Sitter solution for the cosmological dynamics if a suitable form of potential function $V(\phi)$ (or equivalently, $f(R)$) is selected. But, no matter what a potential function  is selected, there is always an early time repeller solution corresponding to a radiation dominated universe.  However, it should be stressed that  a long enough period of matter domination, in which structure is supposed to be formed,  is crucial for a successful cosmology  and   a lack of a matter dominated era could make an accelerating model impracticable.  Therefore, it may be an alternative to dark energy for the late-time acceleration, only if a matter dominated era  can appear naturally between the radiation dominated era and the late-time accelerating phase.  On the other hand, we presented a set of equations for scalar linear perturbations and considered the Newtonian limit for static spherically symmetric solutions. It is found  that  the form of potential function $V(\phi)$ is stringently constrained by the solar system test, although the post-Newtonian parameter $\gamma$ is not directly affected by this function. 

Just as in the framework of most higher derivative gravity theories \cite{Simon1990}, it has been shown that ghosts will appear in  most of the non-local gravity models.  In Ref.\cite{Bamba2011}, it is also found that the introduction of a $f(R)$ term into the non-local
action (\ref{action0}) will not contribute to the solution of this problem.  Unfortunately, in the non-local theory investigated here, this problem is still unsolved. One of the three introduced scalar degree of freedom is phantom-like (or a ghost) and the other two are quintessence-like. See Appendix A for the detail discussions.  Introducing a new non-minimally coupled scalar field  into the action of  gravity may alleviate this problem \cite{Zhang2011}.   Although the theory considered here may have some problems in some aspects and there is no direct support from fundamental theories, it is  hard to deny that it appears to provide a small window through which one can look into phenomenological aspects of fundamental theories. 

As is pointed out in Ref. \cite{Nonlocal4},  replacing the operator $\Box^{-1}$ in action (\ref{action0}) will introduce spurious degrees of freedom, because even when $f(\Box^{-1}R)$ is constant, there exist nontrivial homogeneous and isotropic vacuum solutions of the Einstein solution in the scalar-tensor representation of the theory. However, we find that the two cosmological solutions considered in the work are not artifacts of the scalar-tensor theory  (\ref{eq:action4}), but should be two real solutions of the original theory  (\ref{eq:OrginAction}), although we are now not sure that whether the field content of the original theory and the triple-scalar-tensor representation are the same.  This may deserve thorough  investigation in the future.  

\appendix

\section{Einstein frame and ghost problem}
Let us perform a conformal transformation to the Einstein frame
\begin{equation}
g_{\mu\nu}\rightarrow \tilde{g}_{\mu\nu}= \Omega^2 g_{\mu\nu},
\end{equation}
hence 
\begin{equation}
R\rightarrow \tilde{R}=\Omega^{-2}\left[R-6(\Box\ln \Omega+g^{\mu\nu}\nabla_{\mu}\ln\Omega\nabla\ln\Omega)\right],
\end{equation}
with $\Omega^2=(1+\psi+\xi\phi)^{-1}\equiv e^{\Phi}$. Therefore,  for the vacuum case, action (\ref{eq:action4}) can be rewritten as
\begin{eqnarray}
S&=&\int d^4x \frac{\sqrt{-g}}{2\kappa}\left\{ R-\frac{3}{2}\nabla^{\lambda}\Phi\nabla_{\lambda}\Phi-e^{\Phi}
\nabla^{\lambda}\xi\nabla_{\lambda}\Phi\right.\nonumber\\
&-&\left.e^{2\Phi}\phi\nabla^{\lambda}\xi\nabla_{\lambda}\xi -e^{2\Phi}\xi\nabla^{\lambda}\xi\nabla_{\lambda}\phi- e^{2\Phi}\xi V(\phi)\right\}
\end{eqnarray}
Defining $M$  as the coefficient matrix of  the kinetic terms 
\begin{equation}
M=\left(\begin{matrix}
 \frac{3}{2} & \frac{1}{2}e^{\Phi} & 0  \\ 
  \frac{1}{2}e^{\Phi} & e^{2\Phi}\phi  &  \frac{1}{2}\xi e^{2\Phi} \\ 
0 & -\frac{1}{2}\xi e^{2\Phi} &  0
\end{matrix} \right)
\end{equation}
and $\lambda$ as its eigenvalue. Since $M$ is real  $3\times 3$ symmetric matrix, there must exist three real eigenvalues for the kinetic terms, which are determined by the condition $\det (M-I \lambda)=0$. In order to avoid the ghost degree of freedom,  all the three values of  $\lambda$ must be positive. However,  the product of the three values is determined by $\det M=-\frac{3}{8}e^{4\Phi}\xi^2$, which is always negative and the sum of them is $\frac{3}{2}+e^{2 \Phi } \phi$ which is positive. This means that one of the three eigenvalues of the coefficient matrix $M$ is negative  and  the other two are positive.  

\section*{Acknowledgments}

We thank Ying-Li Zhang and Chao-Jun Feng for helpful discussions.
This work is supported in part by Innovation Program of Shanghai Municipal Education Commission under Grant No. 09YZ148 
and Shanghai Education Commission Foundation for Excellent Young High Education Teacher of China under Grant No. SXY-08009.


\begin{thebibliography}{99}
\bibitem{Copeland:2006ej}
E. J. Copeland, M. Sami and S. Tsujikawa,{ Int. J. Mod. Phys. D} {\bf 15}, 1753 (2006).

\bibitem{Li2011}
M. Li, X.-D. Li, S. Wang, Y. Wang, arXiv:1103.5870v3 [astro-ph.CO].

\bibitem{Witten:2002}
E. Witten, arXiv:hep-ph/0002297.

\bibitem{Giddings2006}
S. B. Giddings, Phys. Rev. D74, 106005 (2006).

\bibitem{Deser2007}
S. Deser and R. P. Woodard, Phys. Rev. Lett. 99, 111301 (2007).

\bibitem{Nojiri2008}
S. Nojiri and S. D. Odintsov, Phys. Lett. B 659, 821 (2008).

\bibitem{Koivisto2008} 
T.  Koivisto Phys. Rev. D 77, 123513 (2008).

\bibitem{Koivisto2008b}
T.  Koivisto Phys. Rev. D 78, 123505 (2008).

\bibitem{Nonlocal} 
G. Calcagni, M. Montobbio and G. Nardelli, Phys. Lett. B 662, 285 (2008).
\bibitem{Nonlocal2}
S. Jhingan \textit{et al}, Phys. Lett. B 663, 424 (2008).
\bibitem{Nonlocal3}
S. Capozziello, E. Elizalde, S. Nojiri and S. D. Odintsov, Phys. Lett. B 671, 193 (2009).
\bibitem{Nonlocal4}
N. A. Koshelev, Grav. Cosmol. 15, 220 (2009).
\bibitem{Nonlocal5}
S. Nesseris,  A. Mazumdar, Phys. Rev. D 79, 104006 (2009).
\bibitem{Nonlocal6}
C. Deffayet,  R. P. Woodard,  JCAP 0908, 023 (2009).
\bibitem{Nonlocal7}
T. Biswas, T. Koivisto and A. Mazumdar, JCAP 1011, 008 (2010).
\bibitem{Calcagni2010}
G. Calcagni, G. Nardelli, Phys. Rev. D 82, 123518 (2010).
\bibitem{Nonlocal8}
N. Barnaby, Nucl. Phys. B 845, 1 (2011).
\bibitem{Nonlocal9}
S.  Nojiri, S. D. Odintsov, M. Sasaki, Y.-L.  Zhang,  Phys. Lett. B 696, 278 (2011). 

\bibitem{Will2005}
C. M. Will, Living Rev. Relativity 9, 3(2005).

\bibitem{Simon1990}
 J. Z. Simon, Phys. Rev. D 41, 3720 (1990).
 
 \bibitem{Bamba2011}
 K. Bamba, S. Nojiri, S. D. Odintsov and M. Sasaki, arXiv:1104.2692 [gr-qc].
 
 \bibitem{Zhang2011}
 Y. -L. Zhang,  private communication, 2011. 
\end{thebibliography}
\end{document}